\documentstyle[12pt]{article}

\begin{document}

\title{Nonuniversality in short-time critical dynamics}
\author{C. S. Sim\~{o}es and J. R. Drugowich de Fel\'{i}cio \and Departamento de
F\'{i}sica e Matem\'{a}tica \and Faculdade de Filosofia, Ci\^{e}ncias e
Letras de Ribeir\~{a}o Preto - USP \and Av. Bandeirantes, 3900 - \and %
14040-901 Ribeir\~{a}o Preto, S.P., Brazil}
\maketitle

\begin{abstract}
We study the behavior of dynamical critical exponents of the two-dimensional
Ising model with a line of defects. Simulations done at early time (first
hundred Monte Carlo steps) reveal that the critical exponent $\theta $ of
Janssen {\it et al} depends on the strength of the exchange coupling
constant($J^{\prime }$) of the altered line. On the other hand, our
simulations permit to conclude that the dynamical critical exponent $z$ is
not sensitive to changes in $J^{\prime }$. In addition, we investigate the
possible invariance of the anomalous dimension ($x_{0}$) of the
magnetisation at the begining of the process.
\end{abstract}
\newpage

A lot of results have been recently obtained concerning the critical
dynamical behavior of statistical models at early time. This kind of
investigation was motivated by analytical and numerical results contained in
the papers of Janssen et al.\cite{Janssen} and Huse \cite{Huse} pointing to
a new dynamical critical exponent ($\theta $) describing the now called
''critical initial slip'' phenomena. That exponent describes the raising of
the magnetisation at very short time

\begin{equation}
M(t) \sim m_{0}t^{\theta }  \label{cis}
\end{equation}
when a system initially in random states, with a small magnetisation, is
quenched to the critical temperature and evolves with the dynamics of model
A. Simulations done for the Ising and Potts (q=3) model indicate that the
new exponent $\theta $ does not depend on the kind of dynamics (Metropolis,
heat-bath or Glauber). In addition, generalized Binder's cummulant \cite{Li}
as well as other sample averages \cite{Murilinho}$^{,}$\cite{Jafferson} which
scale as L$^{0}$ have proved to be useful in determining the dynamical
exponent $z$ which relates time and spatial correlation length ($\tau
\propto \xi ^{z}$). The power of this approach was also verified in
nonequilibrium models such as the majority vote model \cite{Maria Augusta}.
In this letter we investigate the behaviour of the dynamical critical
exponents ($z$ and $\theta $) when a marginal operator is present in the
Hamiltonian. Although the presence of a marginal operator has well known
consequences for the static critical behaviour \cite{Kadanoff}, very little
is known for the dynamical one. To the best of our knowledge the only study
on this subject was recently conducted by Li et al. \cite{Ateller} for the
Ashkin-Teller model which interpolates between Ising and $q=4$ Potts model,
exhibiting a continuous varying exponent ($\nu $) for the correlation length
($\xi $). They have also detected a nonuniversal behaviour of the dynamical
exponent $z.$ In the present paper we study the two-dimensional Ising model
with a line of defects (diferent coupling constant between spins along a
given line) to verify the influence of the marginal operator on the
dynamical exponents $\theta $ and $z.$ In fact, we found that along the
altered line the exponent $\theta $ depends on the ratio ($J^{\prime }/J$).
On the contrary, we detected no changes related to the exponent $z$, a
hypothesis that we used to obtain the dependence of the critical exponent $%
\eta $ on the ratio of couplings ($J^{\prime }/J$) is. Finally, we have
investigated the behavior of the anomalous dimension $x_{o}$ of the
magnetisation at the begining of the process. At least within the precision
of our calculations, that value was kept constant no matter the value of the
coupling constant ($J^{\prime }$). In the sequence, we present the model and
the results of the simulations we performed.

The two-dimensional isotropic Ising model undergoes a spontaneous symmetry
breaking \cite{Onsager} at a critical value of the temperature given by $%
T=2J/k\ln 2$ $.$ At the neighborhood of that temperature, the correlation
length diverges as ($T-T_{c}$)$^{-1}$which means that $\nu =1$. In addition,
the magnetization vanishes like ($T-T_{c}$)$^{1/8}$ which leads to the
critical exponent $\beta =1/8.$ At long distances the correlation
energy-energy behaves, at the critical temperature, as $r^{-2x_{\varepsilon
}}$, where $x_{\varepsilon }$ is the anomalous dimension of the energy
satisfying the relation: $x_{\varepsilon }+1/v=2$. So, the introduction of a
different coupling ($J^{\prime }$), along just one line of the lattice ($d=1$%
), fulfils the necessary conditon $x_{op}=d$ $=1$ to obtain nonuniversality 
\cite{Kadanoff}: the presence in the hamiltonian of a operator which scales
as $r^{-d}$. This, in fact, was the result obtained by Bariev \cite{Bariev}
and McCoy and Perk\cite{McCoy}, later rederived by Peschel and Schotte \cite
{Peschel} in the context of quantum chains. After the work of Cardy \cite
{Cardy}, pointing the importance of the conformal invariance in two
dimensional systems, the Ising model with a line of defects was revisited by
Turban \cite{Turban}. In the sequence, many body techniques \cite{Lieb} were
used to investigate its quantum analog \cite{Lula}$^{,}$\cite{Henkel}$^{,}$%
\cite{Henkel2} but after all, this model continues deserving attention \cite
{Affleck}. It is interesting to notice that the ''global'' exponent $\nu $
as well as the critical temperature does not change when the value of the
coupling ($J^{\prime }$) is altered. The same is not true for the exponent ($%
\eta ^{*}=2\beta ^{*}$) of the correlation function {\it along} the line of
defects (at $T=T_{c}$). This ''local '' exponent depends continously on the
value of the coupling ($J$ $^{\prime }$) as

\begin{equation}
\eta ^{*}=(\frac{2}{\pi }\tan ^{-1}K)^{2},  \label{eta}
\end{equation}
where

\[
K=\tanh (J_{1}/kT_{c})/\tanh (J/kT_{c}) 
\]
and $J_{1}$ is the dual of the modified coupling $J^{\prime }$

\[
\exp (-2J_{1}/kT_{c})=\tanh (J^{\prime }/kT). 
\]

Our investigation began by following the evolution of the magnetisation when
samples are sharply prepared with $m_{0}\neq 0$ but small. We measured the
average over samples (10$^{5}$) of the magnetisation for two lines (one pure
, another with defects). In addition, we also calculated the magnetisation
over the entire sample. Simulations were done for lattices with $36\leq
L\leq 72$ and the update followed from heat-bath dynamics. Because our
interest is in the developing of the magnetisation of the lines, we prepared
all the samples with the same initial magnetisation for each line. In
Figures 1.a and 1.b we present the raising of the magnetisation for that
three cases: pure line, line with defects and entire sample, when $J^{\prime
}/J=2$ and $0.5$. We call the attention of the reader for the similarity
between the curves associated to the pure line and to entire sample in both
cases. Even working with small lattices we note that the effect of the line
of defects restricts itself to its neighbourhood. The conclusion is that the
exponent $\theta $ depends on the coupling constant of the defect line (see
Figure 2) as occurs with the static critical index $\eta ^{*}$. Dynamical
universality (same results with heat-bath and Metropolis updating) was
observed in the simulations as found in the pure Ising model \cite{Okano},
although bigger deviations could be seen when $J^{\prime }/J$ $\neq 1$.

Next step is to check the plausible hypothesis that the dynamical critical
exponent $z$ is not changed by the line of defects. As pointed by Li \cite
{Li2}, the short-time behaviour of the magnetisation, of samples that at $%
t=0 $ are totally magnetized ($m=1$), is given by

\begin{equation}
m \sim t^{-\eta /2\nu z}  \label{mag1}
\end{equation}
where $z$ is the dynamical critical exponent, $\eta $ and $\nu $ the known
static indices of the model. If our hypothesis is correct we can obtain the
exponent $\eta ^{*}$ which characterizes the polynomial decay of the
correlation function (at $T=T_{c}$) along the line of defects by comparing
the slope of two straight lines in the log-log plot of magnetisation versus
time (one of them belonging to the line of defects, the other corresponding
to a normal line far enough of the altered one). In Figure 3 is shown the
behaviour of the magnetisation of those lines, for several values of the
coupling constant $J^{\prime }$, whereas in figure 4 are presented estimates
for the exponent $\eta ^{*}$, obtained by the relation

\begin{equation}
\eta ^{*}=\eta \frac{\tan \alpha }{\tan \alpha }^{\prime }  \label{etanum}
\end{equation}
where $\eta =1/4$ is the critical exponent for the pure Ising model. To
obtain that equation we have used that $z$ has ever the same value, no
matter the value of $J^{\prime }$ is. In addition, $\tan \alpha ^{\prime }$%
and $\tan \alpha $ are the slopes of the curves for altered and pure lines.
The results so obtained are in complete agreement with the exact ones
(continuous line) obtained by Bariev \cite{Bariev} and McCoy and Perk\cite
{McCoy}. To obtain the error bars (smaller than the size of points) we
repeated five times each simulation. We observe that our estimates are
independent of the specific value of the dynamical critical exponent $z$.

Another way to get the same conclusion is to calculate the evolution of the
second moment of the magnetisation of samples which at $t=0$ satisfy the
conditions : $m=0$ and $\xi =0$. Scaling arguments \cite{Li} asserts that $%
m^{2}$ should behave as

\begin{equation}
<m^{2}>\sim t^{(d-\eta /\nu )/z}  \label{scaling2}
\end{equation}
where $d$ is the dimension of the object for which we are calculating that
quantity ($d=1$ when we are interested in lines). By saving the results for
several values of $J^{\prime }/J$ we have obtained the Figure 5 which shows
polynomial behavior with distinct exponents. Finally, we can estimate $\eta
^{*}$ from the equation

\begin{equation}
\eta ^{*}=1-(1-\eta )\frac{\tan \phi ^{\prime }}{\tan \phi }  \label{etanum2}
\end{equation}
where $\tan \phi ^{\prime }$ ($\tan \phi $) are the slopes of the straight
lines in the figure, corresponding to the line with (without) defects.
Results for $\eta ^{*}$ obtained by this procedure are plotted in Figure 6.
Once again we see that the hypothesis of constant $z$ works well, at least
when $J^{\prime }/J<1.$ The disagreement observed for higher values of the
ratio $J^{\prime }/J$ we atribute to the influence of the strong coupling
over the fluctuations of the magnetisation of the pure line.

Finally, we check the behavior of the anomalous dimension $x_{0}$ of the
magnetisation far from equilibrium. As it is now known, the critical initial
slip is a consequence of the difference between the anomalous dimension $%
x_{0}$ and the usual scaling dimension $\eta /2\nu $ of the magnetisation 
\cite{Ritschel}. More specifically, we have

\begin{equation}
x_{0}=\eta /2\nu +z\theta  \label{xzero}
\end{equation}
what permits us to obtain $x_{0}$ ,for any value of the ratio ($J^{\prime
}/J $), since we know both: the exponent $\eta $ (Figure 4) and the new
dynamical exponent $\theta $ (Figure 2). As shown in Figure 7, the result is
very approximately equal to 0.53 (corresponding to the pure case) for a
large interval of ($J^{\prime }/J$). We have used the value $2.172\pm 0.006$
,obtained by Grassberger\cite{Grassberger}, for the dynamical critical
exponent $z$.

In summary, in order to understand the effect of a marginal operator over
the dynamical critical exponents we have investigated the short-time
critical dynamics of the Ising model with a line of defects. We have
confirmed nonuniversality for the ''local'' exponent $\theta $ but not for
the global dynamical exponent $z$. The universality of $z$ permitted us to
determine the static nonuniversal index $\eta ^{*}$, of the correlation
function along the line of defects, using just the hundred first steps of
the simulation (free of critical slowing down phenomena). We have also
detected an apparent universal behavior for the anomalous dimension $x_{0}$
of the magnetisation even at the line of defects. Although the numerical
results can be a little bit different, they are qualitatively the same when
we use heat-bath or Metropolis updating. At the present, we are studying
global and local persistence\cite{Majumdar}$^{,}$\cite{Stauffer}$^{,}$\cite
{Schulke} phenomena in this model.

Figure Captions

Figure 1 - Log-log plot of the magnetisation versus time for (a) a pure
line, (b) the line of defects, and (c) the complete square lattice. Samples
(100000) were originally at high temperature but they had a small
magnetisation per site $m_{0}=0.$ $0278.$ The size of the lattice in this
case is L = 72 and the updating was done with heat-bath dynamics. Figure 1.a
shows results for the case $(J^{\prime }/J)$ $=2$ whereas figure 1.b
presents results for $(J^{\prime }/J)=0.5$.

Figure 2 - Dependence of the new critical exponent $\theta $ on the ratio of
couplings $(J^{\prime }/J)$. The continuous curve is just a guide for the
eyes.

Figure 3 - Log-log plot of temporal decay of the magnetisation at the line
of defects for several values of the ratio of couplings ($J^{\prime }/J$).
The slope of those curves depends on the ratio ($J^{\prime }/J$) and give
estimates for $\eta ^{*}$.(Eq. (\ref{etanum})) Heat-bath dynamics was used
to update samples (50000) which were completely magnetized at $t=0$. The
size of lattice was L = 36.

Figure 4 - Estimates for the nonuniversal critical exponent $\eta ^{*}$
(characterizing the decay of the correlation function at the defect line)
obtained by the scaling relation $m\sim t^{-\eta /2\nu z}$,exhibited in the
Figure 3 and Eq. (\ref{etanum}). The continuous line is the exact result for 
$\eta ^{*}$, given by Eq.(\ref{eta}).

Figure 5 - Log-log plot of the squared magnetisation at the defect line for
several values of the ratio of couplings $(J^{\prime }/J)$. Scaling
relations predict that the slope of the lines should be given by $(d-\eta
^{*}/\nu )/z$.

Figure 6 - Estimates for the nonuniversal critical exponent $\eta ^{*}$
obtained with Eq.(\ref{etanum2}). When $J^{\prime }/J$ is greater than 1
results are poor.

Figure 7 - Plot of the anomalous dimension of the magnetisation $%
x_{0}=z\theta +\eta ^{*}/2\nu $ versus the ratio of couplings $(J^{\prime
}/J)$.

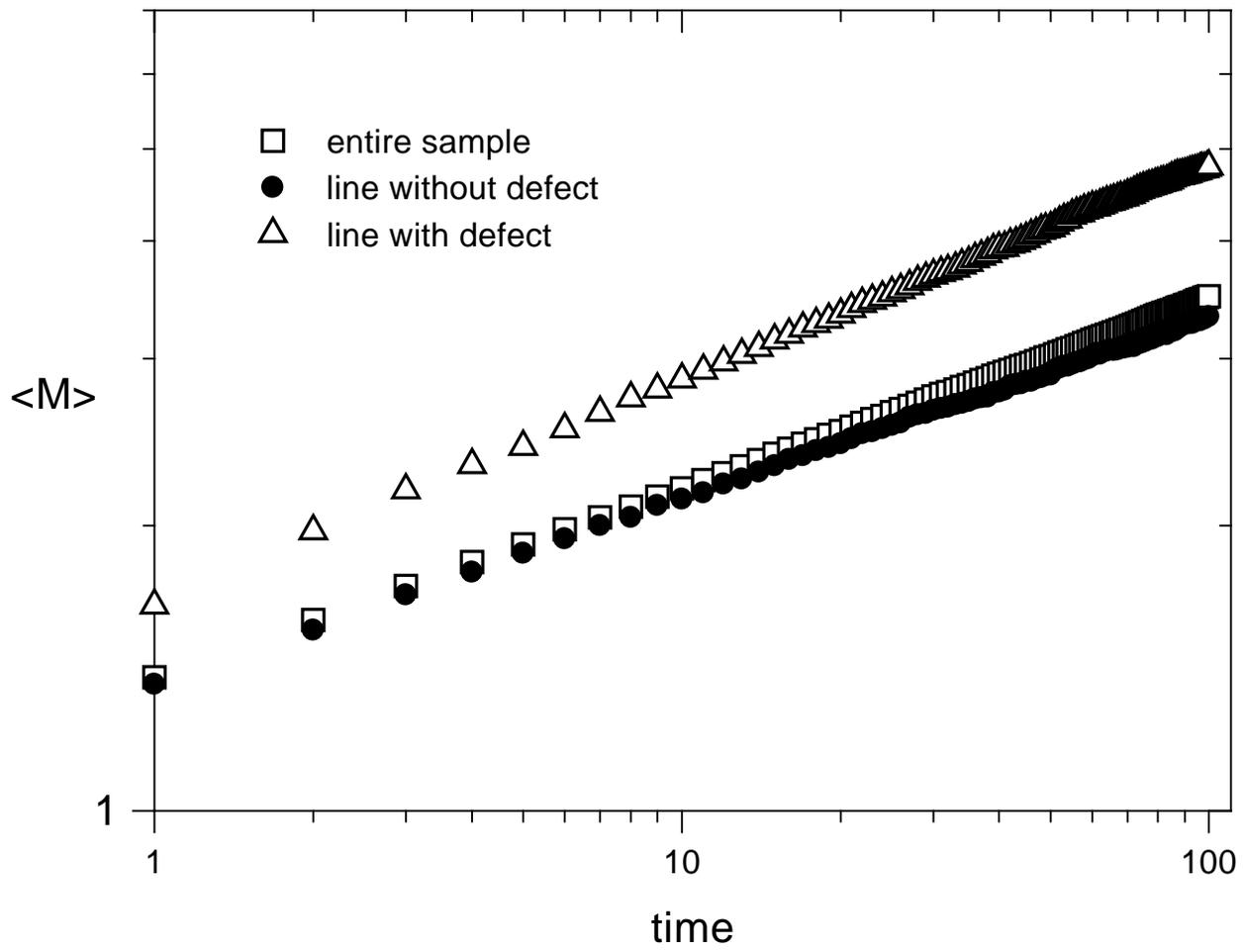
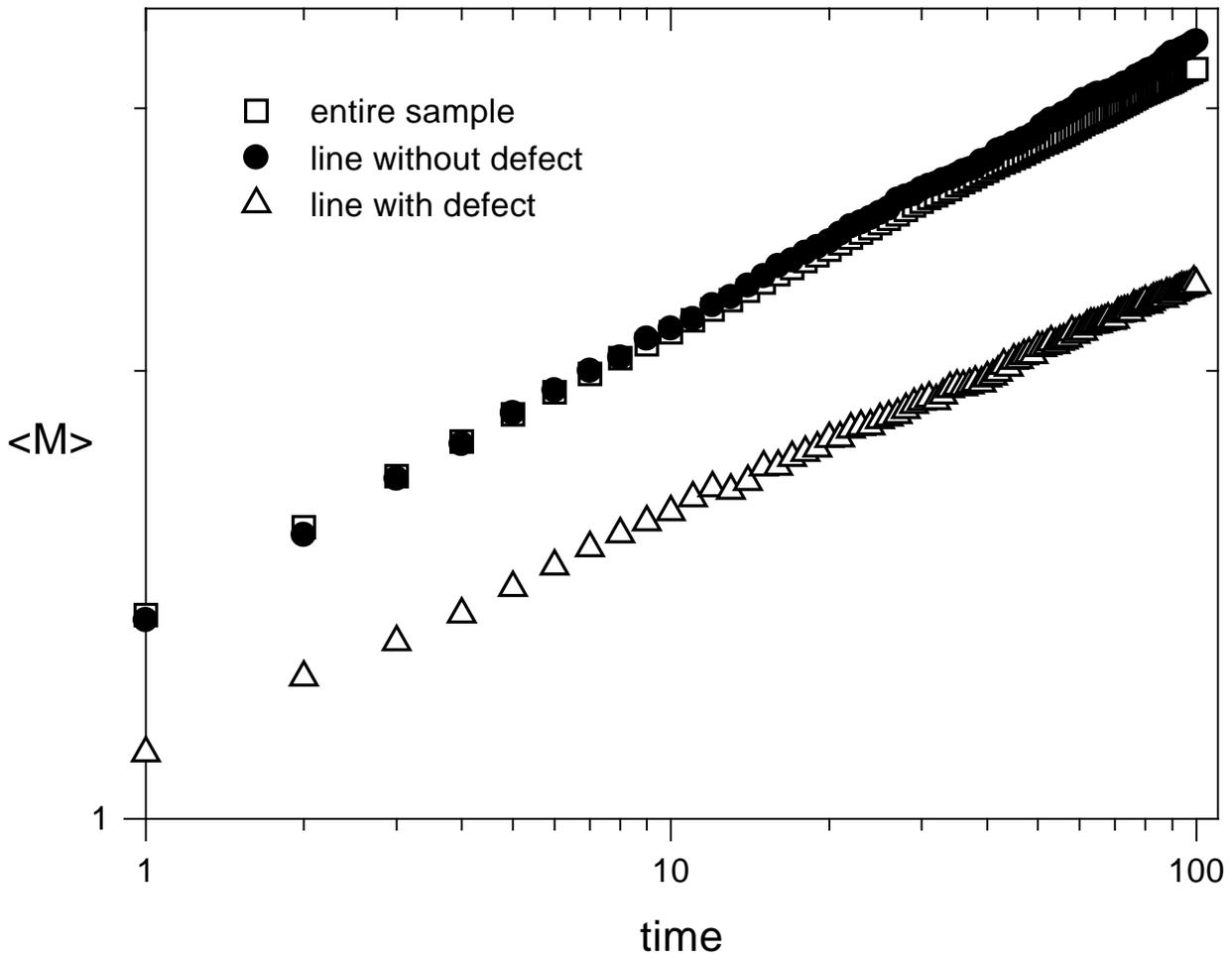
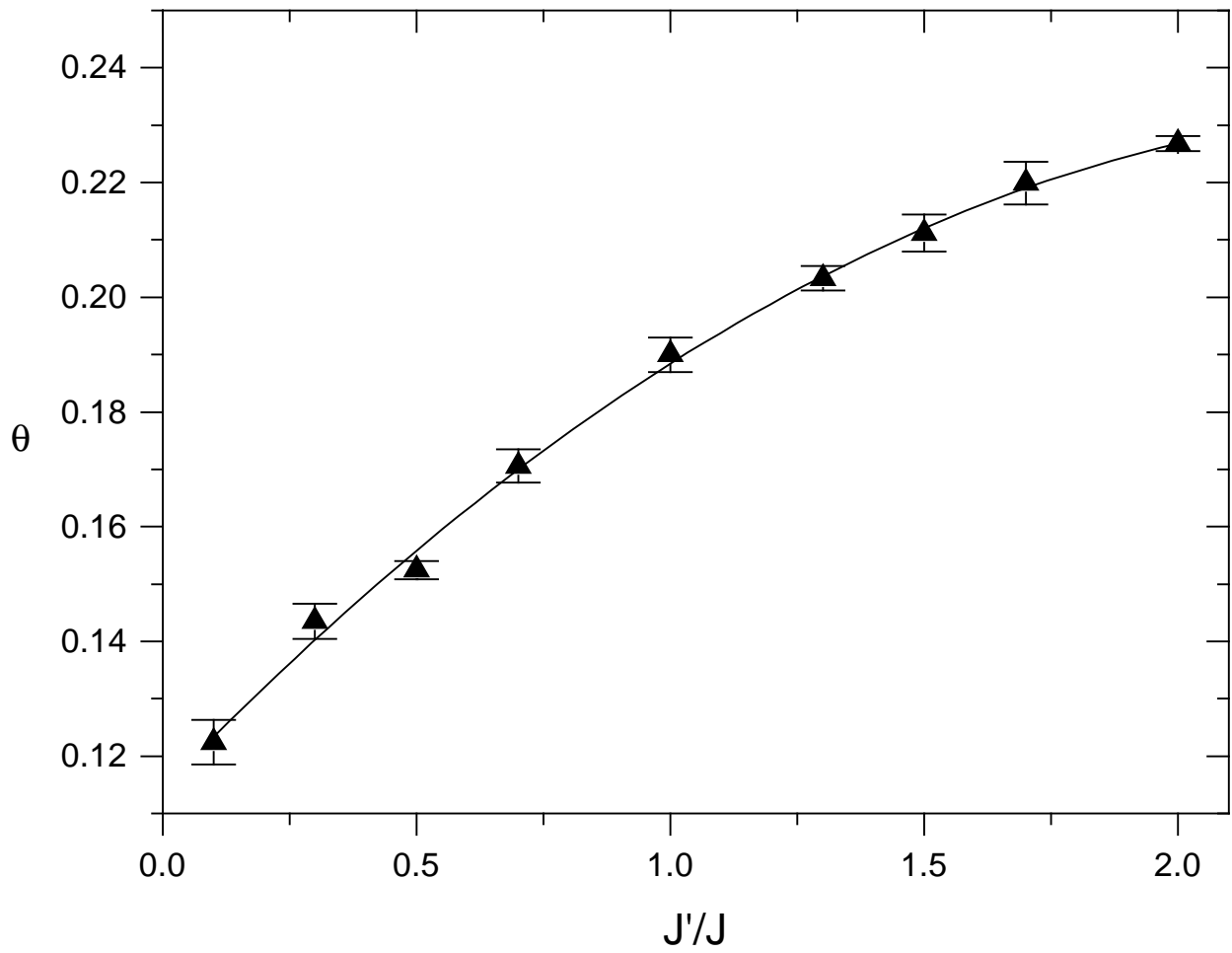
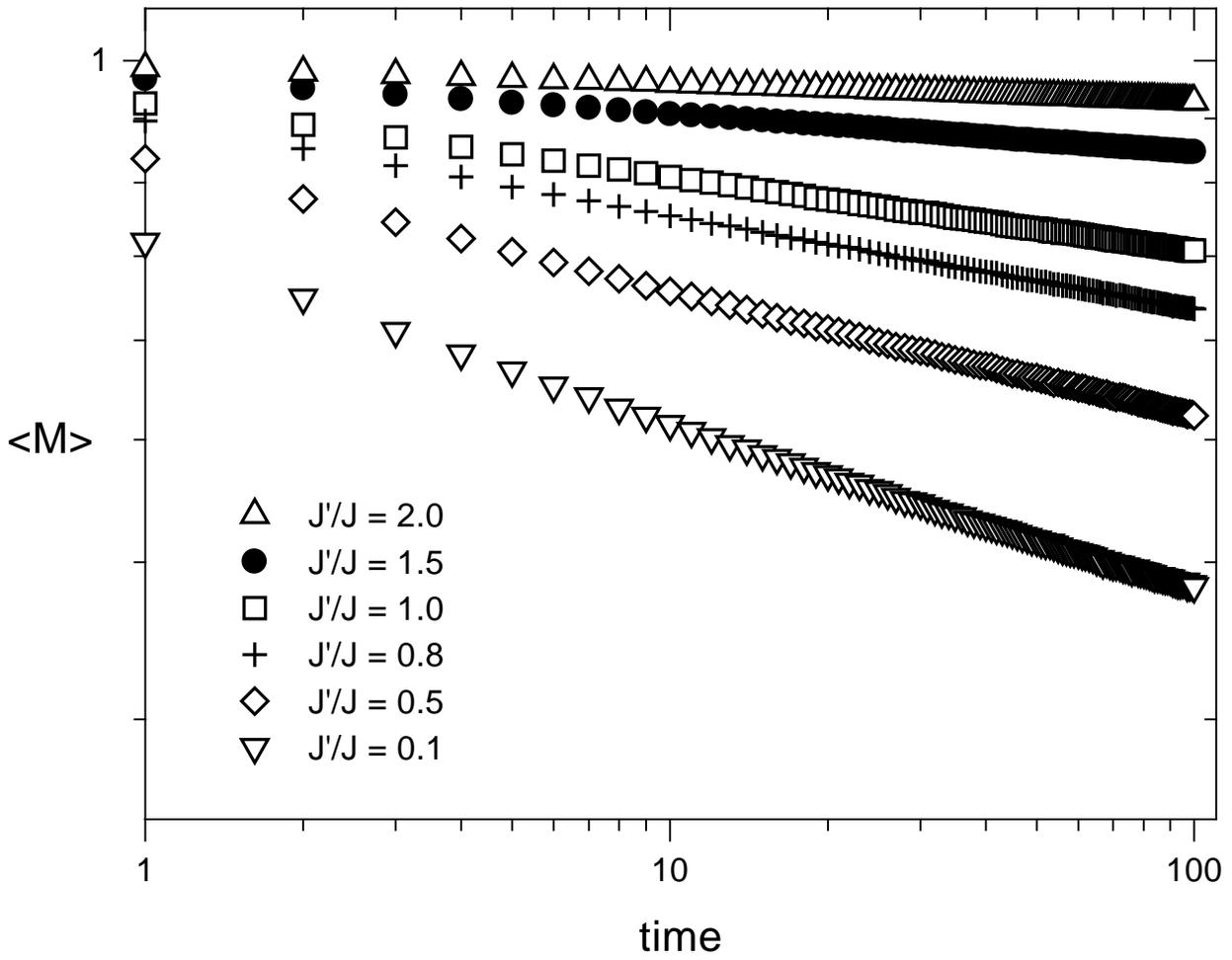
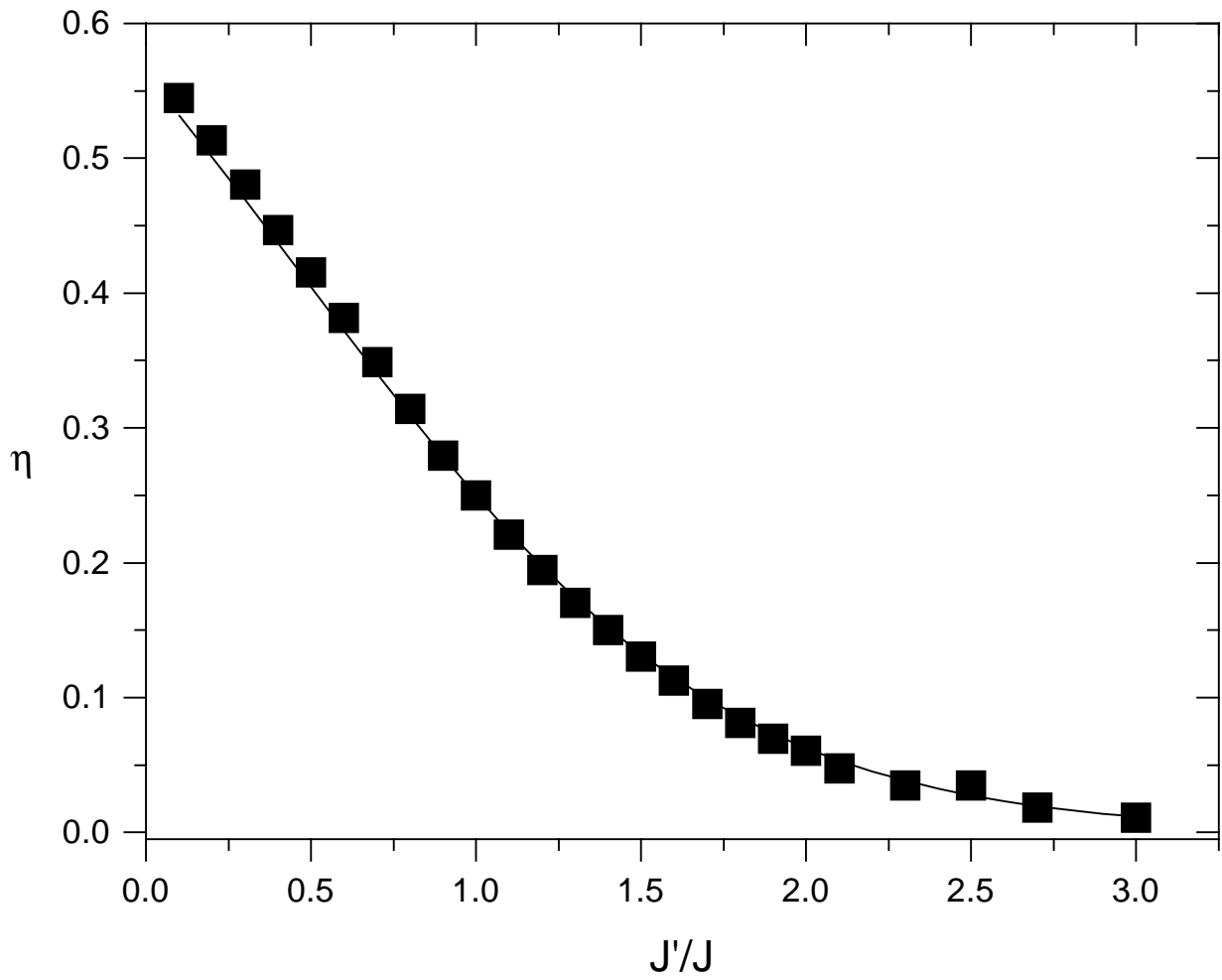
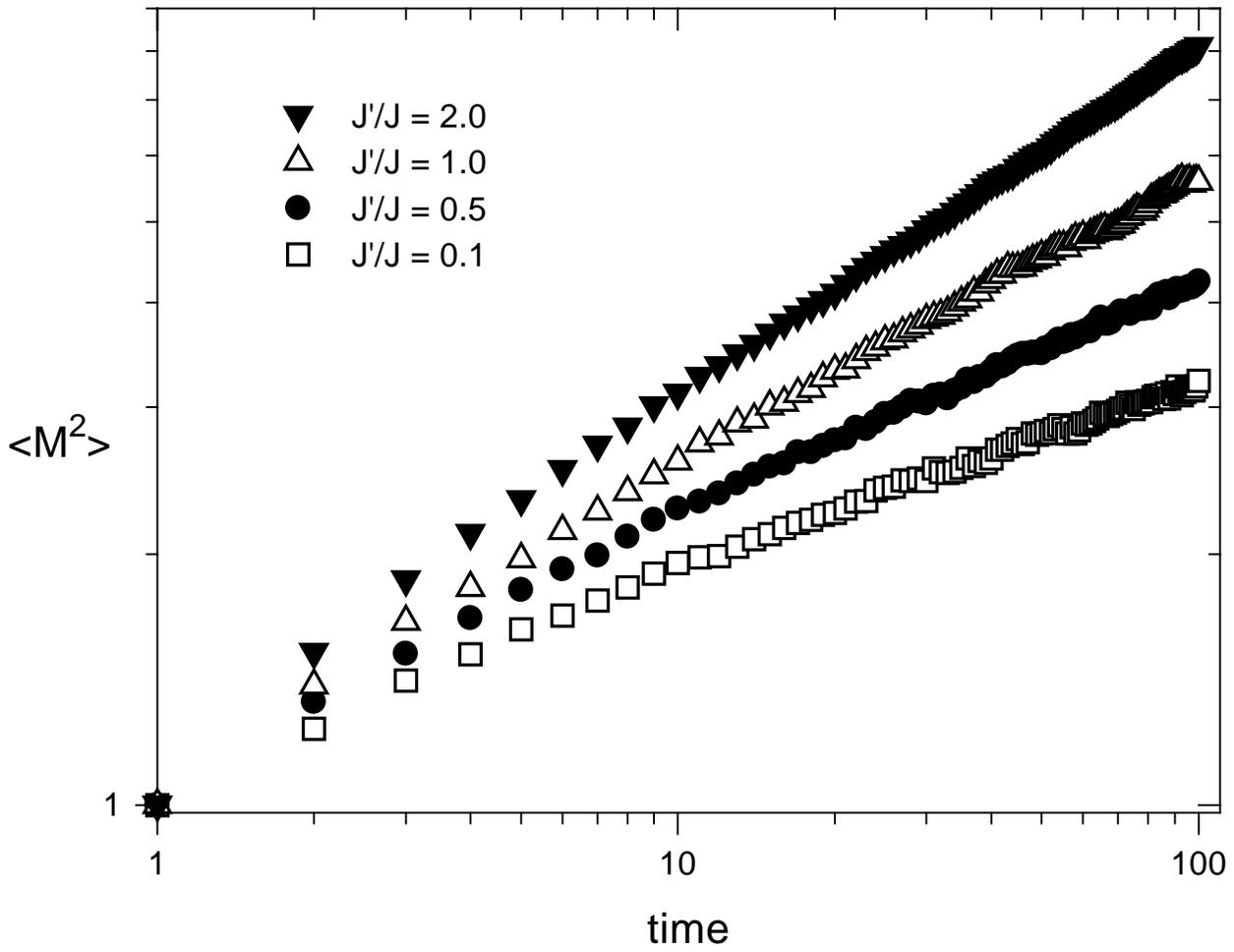
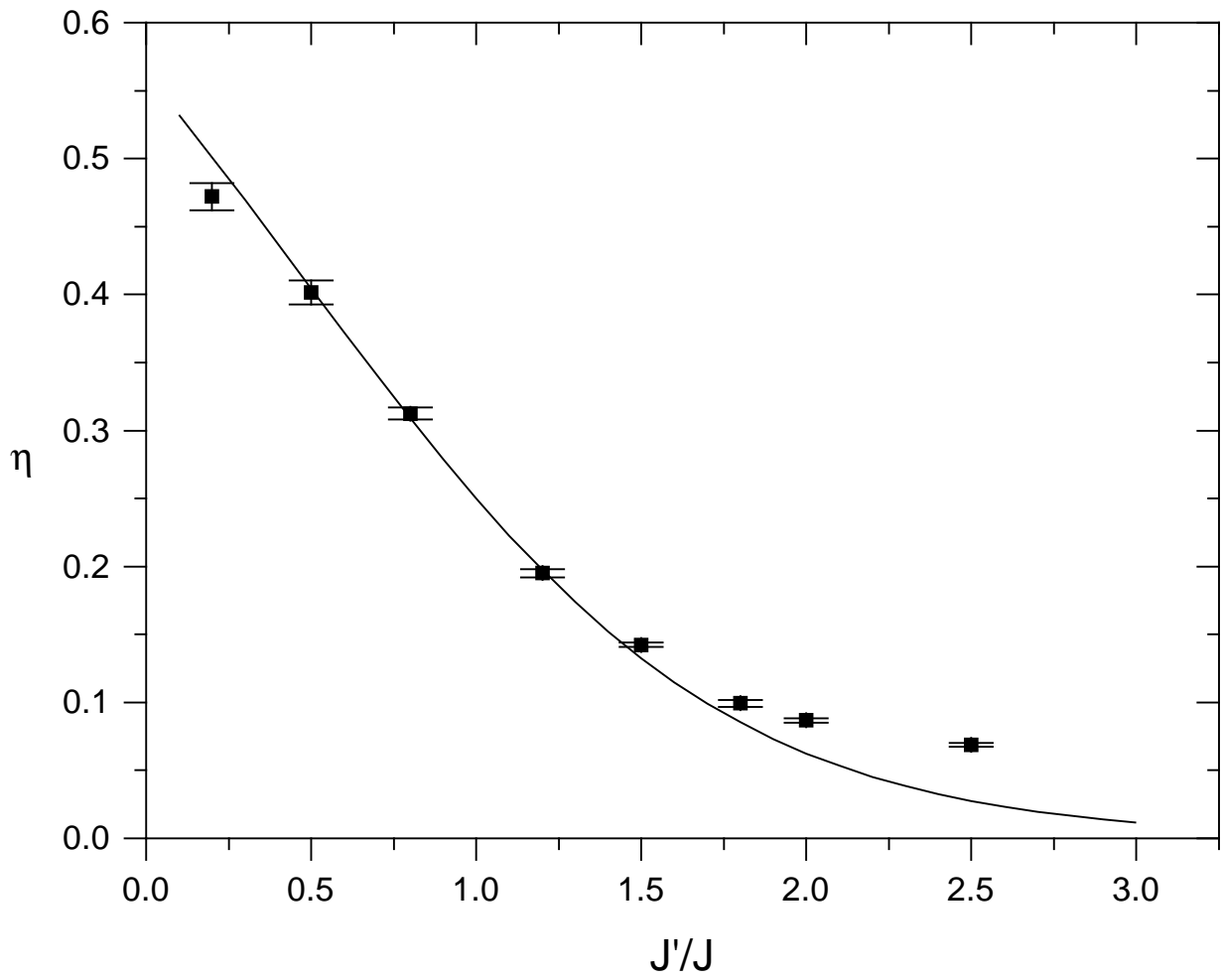
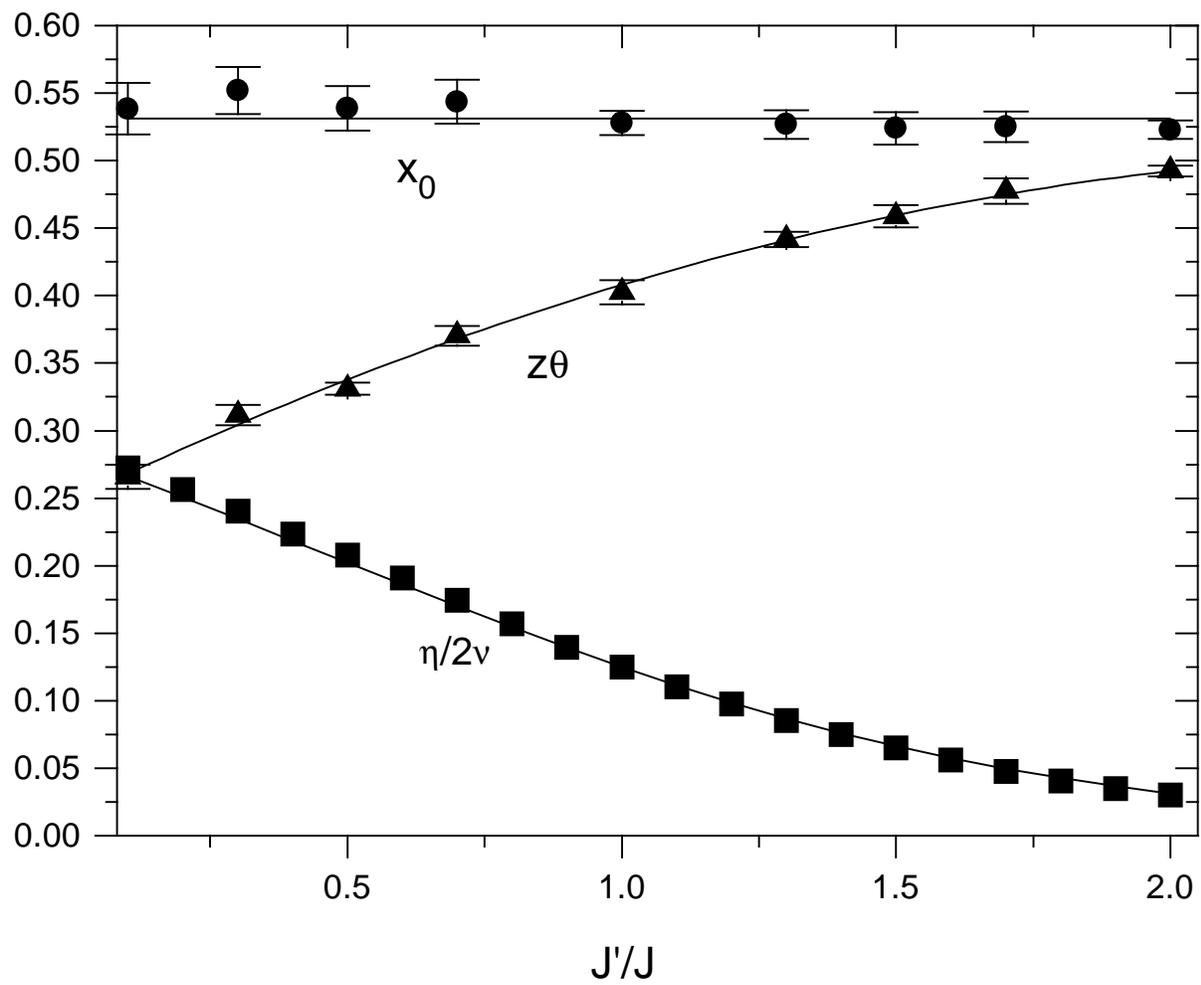
\end{document}